\documentclass[prb,twocolumn]{revtex4-1}
\usepackage{amsmath}
\usepackage{hyperref}
\usepackage{breakurl}
\begin{document}

\title{A Simple Method to Reduce Thermodynamic Derivatives by Computer}

\author{Jacques H. H. Perk}
\email{perk@okstate.edu}
\affiliation{Department of Physics, Oklahoma State University,
Stillwater, OK 74078-3072}

\date{\today}

\begin{abstract}
Studies in thermodynamics often require the reduction of some
first or second order partial derivatives in terms of a smaller
basic set. A simple algorithm to perform such a reduction is
presented here, together with a review of earlier related works.
The algorithm uses Jacobians and is written in Maple language,
but it is easily translated in terms of any other computer
algebra language.
\end{abstract}

\maketitle

\section{Introduction}

About 20 years ago I was teaching a statistical thermodynamics
course using the text by Rumer and Ryvkin.\cite{RR} The students
liked it very much as the book sold new for \$7 only, until MIR
Publishers in Moscow went out of business. From that book I learned
the use of Jacobians for the reduction of partial derivatives in
thermodynamics. This allowed me to create a simple Maple 5.4 algorithm
to systematically do this for me. Recently, I got to teach the
course again using Callen's text,\cite{HBC} which does not use
Jacobians for the reduction, even though a section on it was
included in the first edition.\cite{Cal}

As the Jacobian method is efficient and general, it seems worthwhile
to give a historical review of the method and to present my old
algorithm in Maple 16 format. It is a simple exercise to translate
the code in other formats like Mathematica. At this point I must
mention a Wolfram Demonstrations Project created by Mikhailov\cite{Mik}
about 2009. The Mathematica source code of his CDF file is more or
less equivalent to the first part of my code presented below.

Originally the reduction was done without Jacobians, and large
tables were constructed from which the partial derivatives could be
constructed dividing two entries from these tables, see especially
the works by Bridgman.\cite{Br1,Br2,Br3} Tobolsky\cite{Tob} discussed
how to amend Bridgman's approach if other thermodynamic variables
like length and tension were involved. Bent\cite{Bent} described an
efficient way to reconstruct the tables from a set of linear relations
among differentials.

The use of Jacobians in the current context goes back to at least
the early 1900s, see the encyclopedia article by Bryan,\cite{Bryan}
where in (78)--(82) he indicated how for a single-component system
all partial derivatives can be reduced in terms of those with two
given independent variables. Two of his equations involve Jacobians
and in footnote 57 he attributes this in part to earlier work by
Rankine as the Jacobians arise as the ratio of the areas of two
infinitesimal parallelograms related by a change of variables. Bryan
also discusses the Jacobian approach in his textbook,\cite{Bryan2}
but he does not go into enough applications to demonstrate its
usefulness.

Shaw\cite{Shaw} used Jacobians to extend the Bridgman
tables,\cite{Br2,Br3} also providing a new big table for the
reduction of second derivatives. His Jacobian approach for first
derivatives reproduces the results of Bridgman and he organized
these also in one huge table. But unless one uses his tables regularly
and is very familiar with them, one may prefer to derive the results
oneself using the Jacobian method, as advocated by
Crawford\cite{Cr1} and Pinkerton.\cite{Pink} The efficiency of
the Jacobian approach has been claimed by various other
authors\cite{Cr1,Pink,Hak,SA,GA,CR} and in textbooks like the ones
of Tribus\cite{Tri}, Landau and Lifshitz\cite{LL}, and Jaynes.\cite{Jaynes}
The case of three or more independent variables has also been
addressed in the Jacobian approach.\cite{Cr2,Cr3,MM,Car}

Finally, some authors have sought to implement the reduction method
based on Jacobians using computers. In the mid 1980s
Farah and Missen\cite{FM1,FM2} used muMath-83, a list-processing
language like LISP. Very recently Cooper\cite{JBC} described an
implementation using Mathematica that seems more complicated than
needed, also invoking Gr\"obner bases. The only simple implementation
seems to be the Mathematica app created by Mikhailov\cite{Mik}
mentioned above. The code presented below does more, however, as it
also addresses second derivatives and can be generalized in various
directions including third or higher derivatives or more independent
variables.

\section{Jacobians}

Even though the properties of Jacobians have been reviewed in
several of the above citations, it may be good to summarize them
here. A Jacobian is defined as the determinant of first derivatives:
\begin{equation}
D\equiv\frac{\partial(x_1,x_2,\ldots,x_n)}
{\partial(y_1,y_2,\ldots,y_n)}\equiv
\begin{vmatrix}
\frac{\partial x_1}{\partial y_1}&
\frac{\partial x_1}{\partial y_2}&\cdots&
\frac{\partial x_1}{\partial y_{n_{\vphantom{y}}}}\\
\frac{\partial x_2}{\partial y_1}&
\frac{\partial x_2}{\partial y_2}&\cdots&
\frac{\partial x_1}{\partial y_n}\\
\vdots&\vdots&\ddots&\vdots\\
\frac{\partial x_n}{\partial y_1}&
\frac{\partial x_n}{\partial y_2}&\cdots&
\frac{\partial x_n}{\partial y_n}
\end{vmatrix}.
\end{equation}
Jacobians satisfy the following fundamental properties:
\begin{enumerate}
\item{\textit{Antisymmetry}: A Jacobian is fully antisymmetric
under the exchange of rows or columns, e.g.,
\begin{eqnarray}
&&\displaystyle{
\frac{\partial(x_1,\ldots,x_j,x_{j+1},\ldots,x_n)}
{\partial(y_1,\ldots,y_n)} }\nonumber\\
&&\displaystyle{\qquad=
-\frac{\partial(x_1,\ldots,x_{j+1},x_j,\ldots,x_n)}
{\partial(y_1,\ldots,y_n)}},\nonumber
\end{eqnarray}
\begin{eqnarray}
&&\displaystyle{
\frac{\partial(x_1,\ldots,x_n)}
{\partial(y_1,\ldots,y_j,y_{j+1},\ldots,y_n)}} \nonumber\\
&&\displaystyle{\qquad=
-\frac{\partial(x_1,\ldots,x_n)}
{\partial(y_1,\ldots,y_{j+1},y_j,\ldots,y_n)}},
\end{eqnarray}
which together generate the full antisymmetry.}
\item{\textit{Multiplication}:
\begin{equation}
\frac{\partial(x_1,\ldots,x_n)}
{\partial(y_1,\ldots,y_n)}=
\frac{\partial(x_1,\ldots,x_n)}
{\partial(z_1,\ldots,z_n)}\;
\frac{\partial(z_1,\ldots,z_n)}
{\partial(y_1,\ldots,y_n)},
\end{equation}
which is a direct consequence of the chain rule
$\frac{\partial x_i}{\partial y_j}=\sum_{k=1}^n
\frac{\partial x_i}{\partial z_k}\frac{\partial z_k}{\partial y_j}$
and $\det AB=\det A\det B$.}
\item{\textit{Reciprocity}:
\begin{equation}
\frac{\partial(x_1,\ldots,x_n)}
{\partial(y_1,\ldots,y_n)}\;
\frac{\partial(y_1,\ldots,y_n)}
{\partial(x_1,\ldots,x_n)}=1,
\end{equation}
following from multiplication and $\frac{\partial x_i}{\partial x_j}=
\delta_{ij}$, (which is 1 if $i=j$, and 0 if $i\ne j$).}
\item{\textit{Reduction}: Again using $\frac{\partial x_i}{\partial x_j}=
\delta_{ij}$, we have
\begin{eqnarray}
&&\frac{\partial(y_1,\ldots,y_m,x_{m+1},\ldots,x_n)}
{\partial(x_1,\ldots,x_m,x_{m+1},\ldots,x_n)} \nonumber\\
&&\qquad=
\left.\frac{\partial(y_1,\ldots,y_m)}
{\partial(x_1,\ldots,x_m)}\right|_{x_{m+1},\ldots,x_n=\rm const}.
\end{eqnarray}
For example,
\begin{eqnarray}
&&\frac{\partial(y_1,x_2,\ldots,x_n)}
{\partial(x_1,x_2,\ldots,x_n)} \nonumber\\
&&\qquad=
\begin{vmatrix}
\frac{\partial y_1}{\partial x_1}&
\frac{\partial y_1}{\partial x_2}&\cdots&
\frac{\partial y_1}{\partial x_{n_{\vphantom{y}}}}\cr
0&1&\cdots&0\cr
\vdots&\vdots&\ddots&\vdots\cr
0&0&\cdots&1
\end{vmatrix}
=\left(\frac{\partial y_1}
{\partial x_1}\right)_{x_2,\ldots,x_n}.
\end{eqnarray}
}
\end{enumerate}
These properties mean that we can treat the
$\partial(x_1,x_2,\ldots,x_n)$ formally as if they are numbers.
More precisely, we can define
\begin{equation}
J(x_1,x_2,\ldots,x_n)
\equiv\frac{\partial(x_1,x_2,\ldots,x_n)}
{\partial(b_1,b_2,\ldots,b_n)},
\end{equation}
for a given preferred basic set of independent variables
$b_1,b_2,\ldots,b_n$. Then,
\begin{equation}
\frac{\partial(x_1,x_2,\ldots,x_n)}
{\partial(y_1,y_2,\ldots,y_n)}=
\frac{J(x_1,x_2,\ldots,x_n)}
{J(y_1,y_2,\ldots,y_n)},
\end{equation}
which also explains how the original work of
Bridgman\cite{Br1,Br2,Br3} works.\cite{Shaw}

The above properties imply also the well-known
minus-one and plus-one rules often used in thermodynamics:
\begin{enumerate}
\setcounter{enumi}{4}
\item{\textit{Minus-one rule}:
If $f(x,y,z)=0$, then
\begin{eqnarray}
&{\displaystyle\left(\frac{\partial x}{\partial y}\right)_{\!z}
\left(\frac{\partial y}{\partial z}\right)_{\!x}
\left(\frac{\partial z}{\partial x}\right)_{\!y}}&=\;
\frac{\partial(x,z)}{\partial(y,z)}\;
\frac{\partial(y,x)}{\partial(z,x)}\;
\frac{\partial(z,y)}{\partial(x,y)}
\nonumber\\
&&=-\frac{\partial(x,z)}{\partial(y,z)}\;
\frac{\partial(x,y)}{\partial(x,z)}\;
\frac{\partial(y,z)}{\partial(x,y)} \nonumber\\
&&=-1.
\end{eqnarray}
}
\item{\textit{Plus-one rule}:
If $f(x,y,z,u)=0$ and $g(x,y,z,u)=0$, then two of $x,y,z,u$
are fixed if the other two are specified. Then any pair can
be considered as the independent variables, so that
\begin{eqnarray}
&\left(\frac{\partial x}{\partial y}\right)_{\!u}
\left(\frac{\partial y}{\partial z}\right)_{\!u}
\left(\frac{\partial z}{\partial x\vphantom{y}}\right)_{\!u}&=
\frac{\partial(x,u)}{\partial(y,u)}\;
\frac{\partial(y,u)}{\partial(z,u)}\;
\frac{\partial(z,u)}{\partial(x,u)} \nonumber\\
&&=+1.
\end{eqnarray}
}
\end{enumerate}
Finally, Jacobians can also be used as a condition for
exact differentials.
\begin{enumerate}
\setcounter{enumi}{6}
\item{\textit{Exact differential}: The following are equivalent:
\begin{eqnarray}
&\mathrm{d}f=\sum_{j=1}^n f_j\,\mathrm{d}x_j&\quad\textrm{exact differential}
\nonumber\\
&&\Longleftrightarrow\oint\mathrm{d}f=0
\nonumber\\
&&\Longleftrightarrow\int_A^B\mathrm{d}f\quad\textrm{path independent}
\nonumber\\
&&\Longleftrightarrow f_j=\frac{\partial f}{\partial x_j}
\nonumber\\
&&\Longleftrightarrow\frac{\partial f_i}{\partial x_j}=
\frac{\partial f_j}{\partial x_i}
\nonumber\\
&&\Longleftrightarrow\frac{\partial(f_i,x_i)}{\partial(f_j,x_j)}=-1,
\end{eqnarray}
as the last equality implies 
$\frac{\partial(f_i,x_i)}{\partial(x_j,x_i)}=
\frac{\partial(f_j,x_j)}{\partial(x_i,x_j)}$,
which is the line above it. The other equivalences are also
fundamental in junior and higher level mechanics and
electromagnetism courses, to which we can refer.}
\end{enumerate}

\section{Maple Implementation}

Consider a one-component system with a fixed amount of $N$
moles of matter. We want to simplify all first and second
partial derivatives involving the following eight variables:
absolute temperature $T$, entropy $S$, pressure $P$,
volume $V$, and the four thermodynamic potentials\cite{HBC}
$U$, $H$, $F$ (also denoted $A$ at times) and $G=N\mu$
(with $\mu$ the chemical potential),
\begin{eqnarray}
&&U=U(S,V,N),\quad \mathrm{d}U=T \mathrm{d}S-
P \mathrm{d}V+\mu \mathrm{d}N,\nonumber\\
&&H=H(S,P,N),\quad\! \!\mathrm{d}H=T \mathrm{d}S
+V \mathrm{d}P+\mu \mathrm{d}N,\nonumber\\
&&F=F(T,V,N),\quad \mathrm{d}F=-S \mathrm{d}T
-P \mathrm{d}V+\mu \mathrm{d}N,\nonumber\\
&&G=G(T,P,N),\quad \mathrm{d}G=-S \mathrm{d}T
+V \mathrm{d}P+\mu \mathrm{d}N.\quad
\label{UHFG}
\end{eqnarray}

In the Maple code we formally replace $\partial(X,Y)$, with
$X$ and $Y$ chosen from the list of eight variables, by
\texttt{J[n[X],n[Y]]}, where \texttt{n} replaces each variable's
symbol by a number from 1 to 8. From the antisymmetry
of \texttt{J[i,j]} and (\ref{UHFG}) we can now construct the
code for the first derivatives, provided we express four of
the derivatives in terms of the isobaric thermal volume
expansion coefficient $\alpha$, the two specific heats
$c_p$ and $c_v$ and the isothermal compressibility
$\kappa_T$, namely
\begin{eqnarray}
&&\mathtt{ap}=\alpha=\alpha_p=
\frac1V{\Big(\frac{\partial V}{\partial T}\Big)}_P,\nonumber\\
&&\mathtt{cp}=C_p/N=c_p=
\frac{T}{N}{\Big(\frac{\partial S}{\partial T}\Big)}_P,\nonumber\\
&&\mathtt{cv}=C_v/N=c_v=
\frac{T}{N}{\Big(\frac{\partial S}{\partial T}\Big)}_V,\nonumber\\
&&\mathtt{kt}=\kappa=\kappa_T=
-\frac1V{\Big(\frac{\partial V}{\partial P}\Big)}_T.
\label{act}
\end{eqnarray}
It may be noted that in many textbooks $\alpha$ is used for
the linear expansion coefficient and $\beta=3\alpha$ for the
volume one, but this notation is confusing in statistical physics
due to the ubiquitous use of $\beta=1/k_\mathrm{B}T$ .

We can now give the first part of the Maple code:
\begin{verbatim}
n[P]:=1: n[T]:=2: n[V]:=3: n[S]:=4:
n[U]:=5: n[H]:=6: n[F]:=7: n[G]:=8:
J[3,1]:=V*ap*J[2,1]: J[3,2]:=V*kt*J[2,1]:
J[4,2]:=J[3,1]: cv:=cp-T*V*ap^2/kt:
J[4,1]:=cp*J[2,1]/T: J[4,3]:=cv*J[2,3]/T:
ap:=alpha: cp:=N*c[p]: kt:=kappa:
for i to 4 do
 J[5,i]:=T*J[4,i]-P*J[3,i];
 J[6,i]:=T*J[4,i]+V*J[1,i];
 J[7,i]:=-S*J[2,i]-P*J[3,i];
 J[8,i]:=-S*J[2,i]+V*J[1,i]
 end do:
for i from 6 to 8 do for j from 5 to i-1 do
J[i,j]:=(J[i,1]*J[j,3]-J[j,1]*J[i,3])/J[1,3]
end do end do;
for j to 8 do J[j,j]:=0;
 for i to j-1 do J[i,j]:=-J[j,i] end do end do:
d:=(p,q,r)->normal(J[n[p],n[r]]/J[n[q],n[r]]):
\end{verbatim}
Here \texttt{J[i,j]/J[1,3]} is worked out as a Jacobian. We can
now work out ${(\frac{\partial X}{\partial Y})}_Z$ by the
Maple command ``\texttt{d(X,Y,Z);}'', provided $Y$ and $Z$
are not the same, as one cannot vary $Y$ at constant $Y$.
The do loops in the code indicate how the code can be generalized
to the case of three or more independent degrees of freedom.

As an example,  Maple command ``\texttt{d(P,T,U);}'' gives
\begin{equation}
{\Big(\frac{\partial P}{\partial T}\Big)}_U=
-\frac {-PV\alpha_{p}+Nc_{p}}{V( P\kappa_{T}-T\alpha_{p})}.
\end{equation}

So far we used the second derivatives of $G(T,P,N)$, that
can be denoted by $G_{TT}=-S_T$, $G_{TP}=-S_P=V_T$
and $G_{PP}=V_P$ and further expressed through
(\ref{act}). For the Maple code of second derivatives we also
need the four third derivatives of $G(T,P,N)$, given by
$-S_{TT}$, $V_{TT}$, $V_{TP}$ and $V_{PP}$, which can
be expressed through first derivatives of $\alpha$, $c_p$
and $\kappa_T$. We also need \texttt{m}, the inverse of
\texttt{n} satisfying \texttt{n[m[i]]} = \texttt{i}, so that we can
apply the function \texttt{d} defined in the above code.

For the second part of our code we define the Maple
procedure (somewhat like a Module in Mathematica)
as follows:
\begin{verbatim}
m[1]:=P: m[2]:=T: m[3]:=V: m[4]:=S:
m[5]:=U: m[6]:=H: m[7]:=F: m[8]:=G:
SP:=-VT: VTP:=VPT: SPP:=-VPT:
  STP:=-VTT: SPT:=-VTT:
su:=w->simplify(subs(alpha=VT/V,
  kappa=-VP/V,c[p]=T*ST/N,w)):
dd:=proc(x,y,z) local a,ct,k,o; global m,su; 
  o:=su(x);
  ct:=p->su(d(p,y,z)); a:=0;
  for k to 8 do a:=a+diff(o,m[k])*ct(m[k])
    end do;
  a:=a+diff(o,VT)*(VTP*ct(P)+VTT*ct(T))
   +diff(o,VP)*(VPP*ct(P)+VPT*ct(T))
   +diff(o,ST)*(STP*ct(P)+STT*ct(T));
  a:=simplify(a); RETURN(a) end proc:
\end{verbatim}
One can add further code to eliminate $S_{T}$, $S_{TT}$,
etc., in terms of $\alpha$, $c_p$, $\kappa_T$ and their
first derivatives, if one so desires. Now  ``\texttt{dd(X,Y,Z);}''
works out ${\frac{\partial X}{\partial Y})}_Z$ with $X$
a fairly general expression and $Y$ and $Z$ taken
from the list of eight, $T$ through $G$, but with $Y$ and
$Z$ not the same. In general, the results can be quite messy.

To give an example, ``\texttt{dd(d(F,S,P),T,P);}'' gives
\begin{equation} 
{\Big(\frac{\partial^2 F}{\partial T\partial S}\Big)}_P=
-\frac {PS_{T}V_{TT} -P V_{T}S_{TT}+
{S_{T}}^{\!2}-SS_{TT}}{{S_{T}}^{\!2}},
\end{equation}
which result can be processed further, as $S_T=Nc_p/T$,
$V_T=V\alpha_p$, etc. Some of
this processing can be done by making additions to the above
code, but this will depend on possible applications one
has in mind.

\section{Final remarks}

In the previous section we started with $G(T,P,N)$ with $N$ fixed
and with $T$ and $P$ as the pair of independent variables.
Alternatively we could have started with one of three other
thermodynamic potentials, $U(S,V,N)$, $H(S,P,N)$, or $F(T,V,N)$.
This leads to three variations of the previous section with
three different second and four different third derivatives
of the other thermodynamic potential with respect to its own
pair of independent variables.\cite{Shaw}

It is also straightforward to extend the code to higher
derivatives, the way the previous section is set up. Next,
another variation is to replace $P$ by magnetic field $B$
and $V$ by magnetization $M$. One can also treat the
various entropic versions, which are more directly related
to the various statistical-mechanical ensembles as the
role of $S$ and $U$ is interchanged.\cite{HBC}

In order to treat cases of thermodynamic potentials with
three of more independent variables, one must let the $J$
of the previous section depend on three or more integers.
An example is starting with $G(T,P,x,N)$ for a two-component
system with $N_1=xN$ and $N_2=(1-x)N$ moles of the two
components and total amount  $N$ fixed. The code can be
worked out, but it will be much more elaborate.

Finally, it is an easy exercise to implement the above codes
in other computer languages like Mathematica.

%%%%%%%%%%%%%%%%%%%%%%%%%%%%%%%

\section*{Acknowledgments}

This work has been supported in part by the National Science
Foundation under grant No.\ PHY-07-58139.

%%%%%%%%%%%%%%%%%%%%%%%%%%%%%%%


\begin{thebibliography}{0}

\bibitem{RR}
Yu. B. Rumer and M. Sh. Ryvkin,
\textit{Thermodynamics, Statistical Physics, and Kinetics},
translated from the 1977 Russian edition by S. Semyonov,
(Mir Publishers, Moscow, 1980),
ch.~1 and~2.

\bibitem{HBC}
H. B. Callen,
\textit{Thermodynamics and an Introduction to Thermostatistics},
2nd edition (John Wiley \& Sons, New York, NY, 1985).

\bibitem{Cal}
H. B. Callen,
\textit{Thermodynamics ---
an Introduction to the physical theories of thermostatics
and irreversible thermodynamics},
(John Wiley \& Sons, New York, NY, 1960), sec.~7.5.

\bibitem{Mik}
M. D. Mikhailov,
``Derivation of Thermodynamic Derivatives Using Jacobians,"
from the \textit{Wolfram Demonstrations Project}
$<$http://demonstrations.wolfram.com/ DerivationOfThermodynamicDerivativesUsingJacobians$>$.
\bibitem{Br1}
P. W. Bridgman,
``A Complete Collection of Thermodynamic Formulas,"
Phys. Rev. {\bf 3}, 273--281 (1914).

\bibitem{Br2}
P. W. Bridgman,
\textit{A Condensed Collection of Thermodynamic Formulas},
(Harvard Univ. Press, Cambridge, Mass., 1925)

\bibitem{Br3}
P. W. Bridgman,
\textit{The Thermodynamics of Electrical Phenomena in Metals and
A Condensed Collection of Thermodynamic Formulas},
(Dover Publ., New York, NY, 1961).

\bibitem{Tob}
A. Tobolsky,
``A Systematic Method of Obtaining the Relations Between Thermodynamic Derivatives,"
J. Chem. Phys. {\bf 10}, 644--645 (1942).

\bibitem{Bent}
H. A. Bent,
``A Simplified Algebraic Method for Obtaining Thermodynamical Formulas,"
J. Chem. Phys. {\bf 21}, 1408--1409 (1953).

\bibitem{Bryan}
G. H. Bryan,
``Algemeine Grundlegung der Thermodynamik,"
\textit{Encyklop\"adie der mathematischen Wissenschaften
mit Einschluss ihrer Anwendungen},
Band V, 1. Teil,
%F\"unfter Band, Physik, Erster Teil,
edited by A. Sommerfeld (B.G. Teubner, Leipzig, 1903--1921),
pp.71--160, see p. 113.\hfill\break
$<$http://gdz.sub.uni-goettingen.de/dms/load/toc/?PPN=
PPN360709532\&DMDID=DMDLOG\_0001$>$.

\bibitem{Bryan2}
G. H. Bryan,
\textit{Thermodynamics. An introductory treatise dealing mainly
with first principles and their direct applications},
(B.G. Teubner, Leipzig, 1907), ch.~III, pp.20--26.

\bibitem{Shaw}
A. N. Shaw,
``The Derivation of Thermodynamical Relations for a Simple System,"
Phil. Trans. R. Soc. London A {\bf 234} (740), 299--328 (1935).

\bibitem{Cr1}
F. H. Crawford,
``Jacobian Methods in Thermodynamics,"
Am. J. Phys. {\bf 17} (1), 1--5 (1949).

\bibitem{Pink}
R. C. Pinkerton,
``A Jacobian Method for the Rapid Evaluation of Thermodynamic
Derivatives, without the Use of Tables,"
J. Phys. Chem. {\bf 56}, 799--800 (1952).

\bibitem{Hak}
R. W. Hakala,
``A Method for Relating Thermodynamic First Derivatives,"
J. Chem. Ed. {\bf 41} (2), 99-101 (1964)

\bibitem{SA}
C. W. Somerton and \"O. A. Arnas,
``On the use of Jacobians to reduce thermodynamic derivatives",
Int. J. Mech. Eng. Ed. {\bf 13} (1), 9--18  (1984).

\bibitem{GA}
P. D. Gujrati and P. P. Aung,
``Nonequilibrium thermodynamics. III. Generalization of Maxwell,
Clausius-Clapeyron, and response-function relations, and the
Prigogine-Defay ratio for systems in internal equilibrium,''
Phys. Rev. E {\bf 85}, 041129 (2012), see appendix.

\bibitem{CR}
J. B. Cooper and T. Russell,
``On the Mathematics of Thermodynamics,"
arXiv:1102.1540.

\bibitem{Tri}
M. Tribus,
\textit{Thermostatics and Thermodynamics ---
An Introduction to Energy, Information and States of Matter,
with Engineering Applications},
(Van Nostrand, Princeton, NJ, 1961), ch. 9.

\bibitem{LL}
L. D. Landau and E. M. Lifshitz,
\textit{Statistical Physics},
Course of Theoretical Physics, Volume 5,
translated from the Russian by J. B. Sykes and M. J. Kearsley,
2nd edition, (Pergamon Press, Oxford, 1969),
sec.~16.

\bibitem{Jaynes}
E. T. Jaynes,
\textit{Thermodynamics},
ch. 2 (unpublished)
$<$http://bayes.wustl.edu/etj/thermo.html$>$.

\bibitem{Cr2}
F. H. Crawford,
``Thermodynamic Relations in $n$-Variable Systems
in Jacobian Form: Part I, General Theory and Application to Unrestricted Systems,"
Proc. Am. Acad. Arts Sci. {\bf 78}, 165--184 (1950).

\bibitem{Cr3}
F. H. Crawford,
``Thermodynamic Relations in $n$-Variable Systems in Jacobian Form:
Part II, Polyphase Polycomponent Chemical Systems,"
Proc. Am. Acad. Arts Sci. {\bf 83}, 193--220 (1955).

\bibitem{MM}
F. S. Manning and W. P. Manning,
``Derivation of Thermodynamic Relations for Three-Dimensional Systems,"
J. Chem. Phys. {\bf 33}, 1554--1557 (1960)

\bibitem{Car}
B. Carroll,
``On the Use of Jacobians in Thermodynamics,"
J. Chem. Ed. {\bf 42} (4), 218-220 (1965)

\bibitem{FM1}
N. Farah and R. W. Missen,
``The Computer-Derivation of Thermodynamic Equations.
Part I. First and Second Derivatives for Complex Unrestricted Systems," 
Can. J. Chem. Eng. {\bf 64}, 154--157 (1986).

\bibitem{FM2}
N. Farah and R. W. Missen,
``The Computer-Derivation of Thermodynamic Equations.
Part II. First and Second Derivatives for Simple Systems,"
Can. J. Chem. Eng. {\bf 65}, 137--141 (1987).

\bibitem{JBC}
J. B. Cooper,
``Thermodynamical identities --- a systematic approach,''
arXiv:1108.4760.

\end{thebibliography}
\end{document}